\documentclass[aps,pre,onecolumn,superscriptaddress,showpacs]{revtex4}
\usepackage{epsfig,xcolor,graphics}
\usepackage{amsmath, mathrsfs}
\usepackage{amssymb}
\usepackage{siunitx}
\usepackage{natbib}
\graphicspath{{./FIGURES/}}
\begin{document}
\title{Effects of particle compressibility on structural and mechanical properties of compressed soft granular materials}
%How packing fraction can account for particle compressibility in soft granular materials ?
\author{Thi-Lo Vu}
\email{vuthilo@tdtu.edu.vn}
\affiliation{Division of Computational Mathematics and Engineering, Institute for Computational Science, Ton Duc Thang University, Ho Chi Minh City, Vietnam}
\affiliation{Faculty of Civil Engineering, Ton Duc Thang University, Ho Chi Minh City, Vietnam}
\affiliation{LMGC, Universit\'e de Montpellier, CNRS, Montpellier, France}
\author{Saeid Nezamabadi}
\email{saied.nezamabadi@umontpellier.fr}
\affiliation{LMGC, Universit\'e de Montpellier, CNRS, Montpellier, France}
\affiliation{IATE, CIRAD, INRA, Montpellier SupAgro, Universit\'e de Montpellier, F-34060, Montpellier, France}
\author{Serge Mora}
\email{serge.mora@umontpellier.fr}
\affiliation{LMGC, Universit\'e de Montpellier, CNRS, Montpellier, France}

\begin{abstract}
Changes in the mechanical properties of granular materials, induced by variations in the intrinsic compressibility of the particles, are investigated by means of numerical simulations based on the combination of the Finite Element and Contact Dynamics methods. Assemblies of athermal 2D particles are subjected to quasi-static uni-axial compactions up to packing fractions close to $1$. Inspired by the contact mechanics in the Hertz's limit, we show that the effect of the compressibility of the particles both on the global and the local stresses, can be described by considering only the packing fraction of the system. This result, demonstrated in the whole range of accessible packing fractions in case of frictionless particles, remains relevant for moderate inter-particles coefficients of friction. The small discrepancies observed with frictional particles originate from irreversible local reorganizations in the system, the later being facilitated by the compressibility of the particles.
\end{abstract}
\maketitle

%%%%%%%%%%%%%%%%%%%%%%%%%%%%%%%%%
\section{Introduction}

The physical properties of granular solids are mainly determined by their discrete and heterogeneous nature \cite{Hinrichsen2004}, leading to well known general features of these disordered media such as chains forces \cite{Radjai1996}, jamming transition \cite{Majmudar2007}, dilatancy \cite{Kabla2009}, etc. Beyond, specific properties emerge due to the peculiar nature of the particles, as inter-particles interactions (e.g. cohesion \cite{Herminghaus2005,Luding2008} or friction \cite{MiDi2004,Goldenberg2005}), the shape of the particles \cite{Azema2010} or size polydispersity \cite{Shaebani2012,Radjai2015}. Taking into account particles bulk properties in numerical simulations requires either to model the deformations of each particle and to consider their complete constitutive law (e.g. elastic \cite{Vu2019_pre}, plastic \cite{Nezamabadi2017}, brittle \cite{Radjai2018}, visco-plastic, etc.), or to reduce the particles behavior to representative key elements through ad hoc and simplified models. This last strategy is based on approximations that are relevant in the small loads or small particle deformations limits, as it is the case near the jamming transition of a granular material. For instance, granular media made of elastic particles have been described using repulsive potentials associated with hypothetical overlaps between particles \cite{OHern2002}. However, if the deformations of the particles are not infinitesimal anymore as it is the case in many applications (food products, metal powders, colloidal suspensions and clays, etc. \cite{Kabla2012,Lepesant2013,Lorenzo2013,Menut2012}), these kinds of approximation are not accurate enough and a complete description of the individual particles mechanical behavior under deformation becomes necessary.

In this paper, we study the effect of the compressibility of elastic particles on structural and mechanical properties of granular assemblies during uni-axial compressions in a wide range of deformations, from the jamming transition to packing fractions of almost $1$, with particles undergoing finite deformations. Such large deformations require the full continuous description of each particle, associated with an objective constitutive law. For simplicity, we choose for the constitutive law of the particles the isotropic neo-Hookean model. 

The paper is organized as follows. The granular systems investigated in this article are first described in Section \ref{sec : system}, together with details about the numerical method. The variation of the packing fraction as a function of the global deformation of the system is first analyzed in Section \ref{sec : packing fraction}. Section \ref{sec : stress} is devoted to the study of the stress; both at the global and the local scales. Some effects of the compressibility of the particles on the coordination number are also briefly discussed. Section \ref{sec : irreversibility} deals with irreversibility, which is at the origin of the small deviations to the master curve evidenced in the previous section.  Section \ref{sec : conclusion} provides a summary and a discussion.

%%%%%%%%%%%%%%%%%%%%%%%%%%%%%%%%% 

\section{Presentation of the granular system} \label{sec : system}
 \begin{figure}[!h]
\begin{center}
\includegraphics[width=0.5\textwidth]{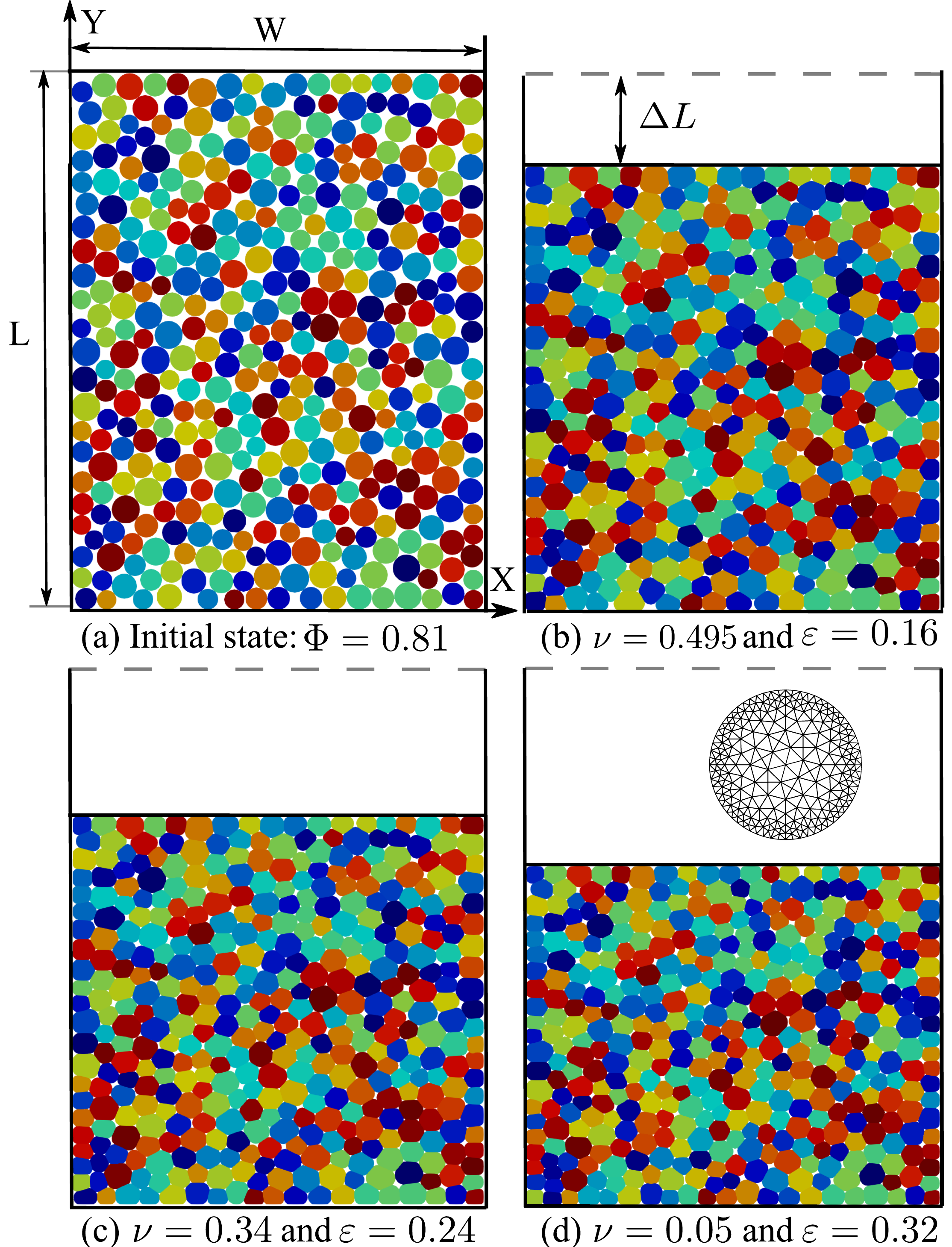}
\end{center}
\caption{
  %A 2d assembly of 400 cylindrical particles subjected to uni-axial compression. The granular material lays in a rectangular box with the width $W = \SI{0.42}{\meter}$ and the initial length $L = \SI{0.51}{\meter}$. The top wall moves at constant velocity $v$ during time $t$. Axis $X$ is set parallel to the mobile wall and axis $Y$ is in the direction of the velocity $v$.  A typical finite element mesh of a particle is shown on the right.
A snapshot of an initial configuration (a) and three snapshots of the compaction of a packing of 400 elastic particles with neo-Hookean behavior for packing fraction of $\Phi=0.97$ and several values of the  Poisson's ratio $\nu$ (b-d). Inter-particles coefficient of friction is $\mu_f=0.4$. Axis $X$ is set parallel to the mobile wall and axis $Y$ is in the direction of the compression. The horizontal dashed lines indicate the initial position of the mobile wall.  A typical finite element mesh of a particle is also shown zoomed in (d). Note that, to achieve the same value of the packing fraction, different values of cumulative compressive strain (see Eq.~(\ref{eqn : cumulative strain})) are needed due to the different compressibilities of the particles. }
\label{fig : mesh}
\end{figure}

In this paper, we consider sets of 400 elastic disks placed in a rectangular box (width $W=\SI{0.42}{\meter}$ along axis $X$ and initial length $L=\SI{0.51}{\meter}$ along axis $Y$) with rigid walls (see  Fig.~\ref{fig : mesh}). One of the two walls aligned along the $X$ direction is mobile inward at a constant and small velocity $v$. The three other walls being fixed, the granular system undergoes an uni-axial compression. The initial configurations are defined by randomly distributing the particle positions in the $XY$ plane of the box prior the mobile wall begins to move. The bulk mechanical properties of the elastic disks are described by the isotropic neo-Hookean constitutive law \cite{Ogden1984} given by the following strain energy density function \cite{Nezamabadi2011,Vu2019_ExpMech}: 
\begin{equation}
  \Psi = \frac{E}{2(1+\nu)}\left(\frac{I_1-3}{2}-\ln J + \frac{\mu}{1-2\nu}\left(\ln J \right)^2 \right)\label{eqn : neo hookean}, 
\end{equation}
with $I_1=\mathrm{Tr} (\mathbf F^T  \mathbf F)$ and $J=\mathrm{det} (\mathbf F)$. $\mathbf F$ denotes the deformation gradient tensor defined as $\mathbf F =  \mathbf I + \mbox{\boldmath $\nabla$} \mathbf u$ ($\mathbf I$ being the second-order identity tensor and $\mathbf u$ the displacement field). $E$ and $\nu$ are the Young's modulus and the Poisson's ratio, respectively. For a given granular system studied in this paper, all the particles have the same Young's modulus and the same Poisson's ratio. The particle diameters are set in the range $[0.01, 0.015]$m, with a size distribution uniform by particle surface fraction; \textit{i.e.}, all size classes have the same surface of particles (crystallization is then inhibited). For simplicity, the contacts are assumed to be frictionless with the walls. Inter-particles friction is assumed to follow Coulomb friction law with the same coefficient of friction $\mu_f$ between all particles of a given granular assembly.

We consider for all simulated systems the Young's modulus and the mass density of the particles to be $E=\SI{0.45}{\mega \pascal}$ and $\rho = 1180$ kg $m^{-3}$. This choice is consistent with particles made of silicon based elastomer \cite{Vu2017,Vu2019_ExpMech}. In order to investigate the effects of the particle compressibility, systems with three different Poisson's ratios are studied : systems made of quasi-incompressible particles with $\nu=0.495$, systems made of compressible particles with $\nu=0.34$, and systems with highly compressible particles with $\nu=0.05$. In addition, different values of the inter-particles coefficient of friction $\mu_f$ (from $0$ to $0.8$) are tested for each of these Poisson's ratios.  

For preparing the initial configuration of these systems, the particles are randomly distributed by keeping packing fractions (defined as the ratio between the cumulative surface of the particle cross sections and the actual box area) constant and equal to $\Phi_0=0.81$. In order to probe the effect of the initial configuration on measured quantities, three simulations with the same physical parameters are performed using three equivalent initial configurations. In the following, error bars reflect the dispersion coming from these initial configurations. The applied velocities $v$ is fixed to $\SI{0.035}{\meter \per \second}$ so that the global deformation fulfilled the requirements for a quasi-static transformation \cite{Vu2019_soft_matter}.

The numerical technique used in this work, is based on coupling of the finite element method (FEM) and contact dynamics (CD) method, and implemented in the LMGC90 code \cite{lmgc90}. The FEM-CD approach has been described and validated in previous papers \cite{Vu2019_pre,Vu2019_soft_matter}.  In order to compute more precisely the contact between particles, and between walls and particles as well as to optimize the computational cost, each particle is discretized using a mesh which is denser at the periphery than in its central region (see Fig.~\ref{fig : mesh}). The number of degrees of freedom of the system is hence reduced. In this manner, each particle is meshed with about 400, 3-nodes triangular elements. 

\section{Packing fraction of the granular system} \label{sec : packing fraction}
\begin{figure}[!h]
\begin{center}
  \includegraphics[width=0.4\textwidth]{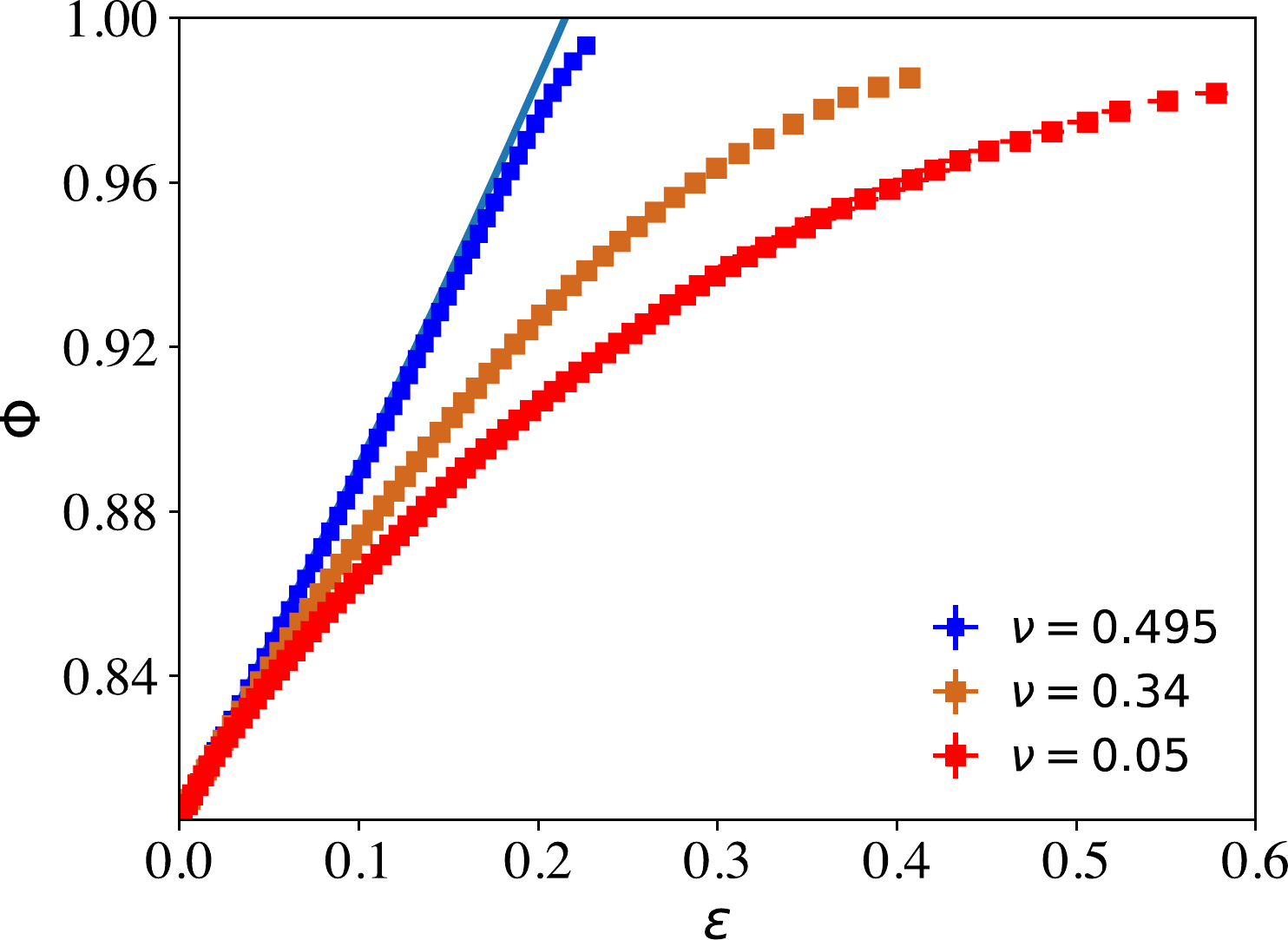}
\end{center}
\caption{Packing fraction $\Phi$ as a function of the cumulative compressive strain for assemblies of particles with $\nu=0.495$ (blue), $\nu=0.34$ (brown) and $\nu=0.05$ (red) with inter-particles coefficient of friction $\mu_f=0.2$. The solid line is the computed packing fraction of assemblies of incompressible particles ($\Phi=\Phi_0\exp \varepsilon$).}
\label{fig : phi}
\end{figure}

In what follows, we consider uni-axial and quasi-static compressions of the granular systems, as described above. In Fig.~\ref{fig : phi}, the packing fraction is plotted as a function of the cumulative compressive strain $\varepsilon$, defined by
\begin{equation}
  \displaystyle{\varepsilon =-\ln (1-\frac{\Delta L}{L}),}
  \label{eqn : cumulative strain}
\end{equation}
for systems with different Poisson's ratios and $\mu_f=0.2$. $\Delta L = vt$ is the displacement of the mobile wall from the beginning of the compaction. For the quasi-incompressible particles ($\nu=0.495$), the cumulative surface of the particles is almost constant, and the packing fraction is well approximated by $\Phi=\Phi_0\exp \varepsilon$ (relation deduced from  Eq.~(\ref{eqn : cumulative strain}) for incompressible particles). 

As expected, the packing fraction of an assembly of compressible particles  for any given value of the imposed deformation, is smaller than the packing fraction of systems with less compressible particles (see movie in supplementary materials, \cite{supplement}). %In addition, the inter-particles coefficient of friction reduces \colorbox{pink}{?} the global compressibility of the granular system.
In the following, mechanical and structural quantitative features of the granular systems will be plotted either as a function of the cumulative compressive strain $\varepsilon$, or as a function of the packing fraction $\Phi$. 

\section{Compressive Stress} \label{sec : stress}
\begin{figure}[!h]
\begin{center}
  \includegraphics[width=0.4\textwidth]{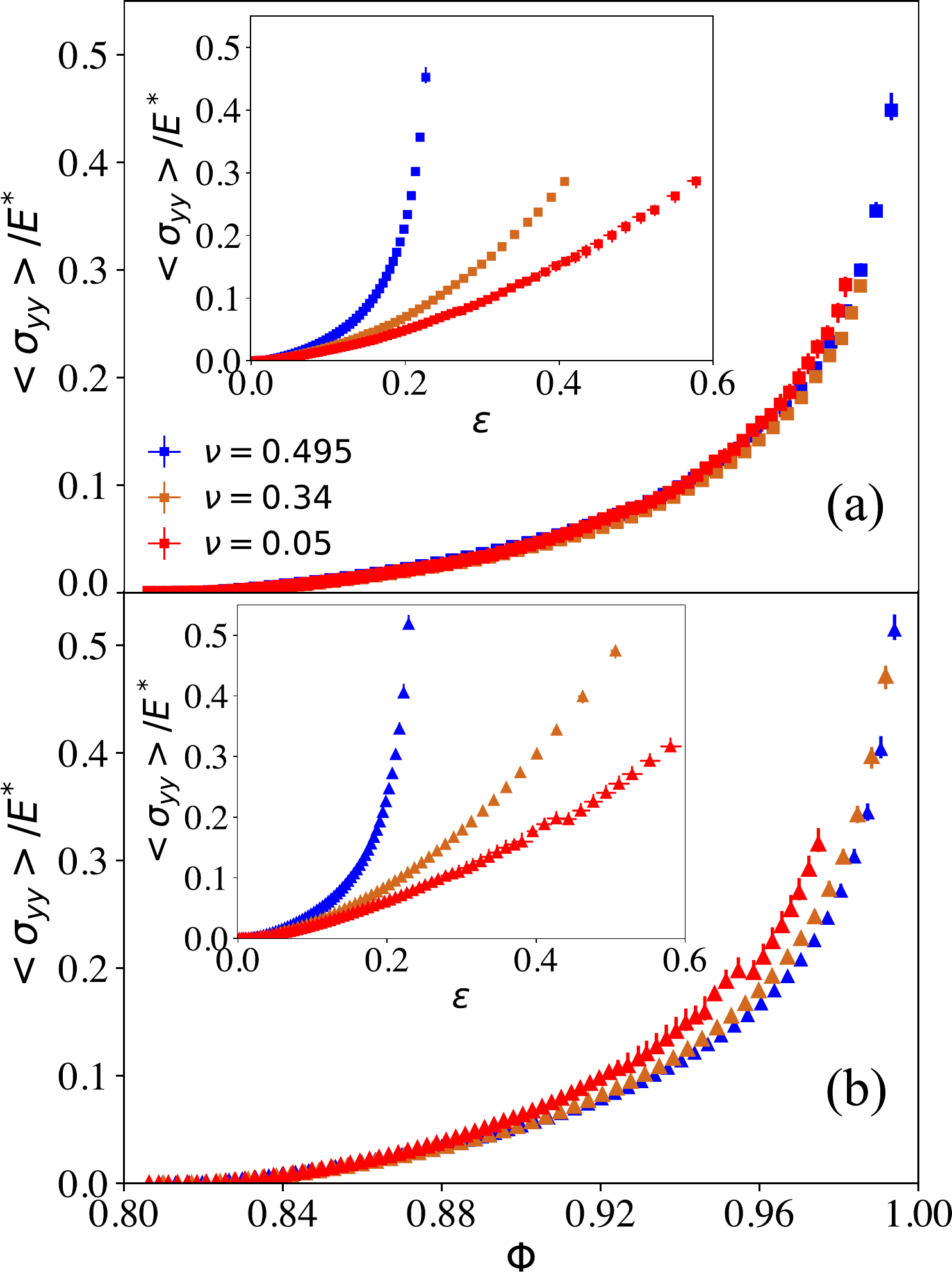}
\end{center}
\caption{Mean value of the normal stress applied to the granular systems, $\langle \sigma_{yy} \rangle$, normalized by the effective Young's modulus $E^*$ of the particles, as a function of the packing fraction $\Phi$ for three different Poisson's ratios of the particles, and for $\mu_f=0.2$ (a) and $\mu_f=0.8$ (b). Inset: Same data in a different representation ($\langle \sigma_{yy} \rangle/E^*$ as a function of $\varepsilon$).}
\label{fig : sigma yy}
\end{figure}

The mean value of the $yy$ component of the Cauchy stress tensor applied to the mobile wall, $\langle \sigma_{yy}  \rangle$, is computed by dividing the contact forces acting on the mobile wall to its length. $\langle \sigma_{yy}  \rangle$ divided by the effective elastic modulus $E^*$ ($E^* = E/(1-\nu^2)$) of the particles is plotted as a function of the cumulative compressive strain $\varepsilon$ in the insets of Fig.~\ref{fig : sigma yy} for systems with $\nu=0.05$, $\nu=0.34$ and $\nu=0.495$, and for $\mu_f=0.2$ and $\mu_f=0.8$. As expected, $\langle \sigma_{yy}  \rangle$ is first equal to zero up to non significant fluctuations, and beyond a certain threshold $\varepsilon_J$ corresponding to the jamming transition, it starts to increase. The corresponding value of the packing fraction will be denoted as $\Phi_J$.  $\varepsilon_J$  depends on $\mu_f$ \cite{Vu2019_soft_matter} but not on $\nu$. For a given cumulative strain, the compressive stress $\langle \sigma_{yy}  \rangle$ is, as expected, smaller as the Poisson's ratio of the particles is close to $0$.    

In the granular systems, the stress is transmitted from one particle to an other one through contact forces. Within the limit of the small deformations of the particles, the contact force (per unit length) between two frictionless elastic disks with the same Poisson's ratio and the same Young's modulus is \cite{Johnson1985}:
\begin{equation}
  F=\frac{\pi}{4}E^*d,
  \label{eqn : Hertz}
  \end{equation}
with $d$ the deflection of each particle. In the FEM-CD simulations, Eq.~(\ref{eqn : Hertz}) has not been used because contact forces are described in the context of finite deformations and then the full resolution of the contact equations has been performed for each contact. However, in order to generalize these properties to the granular systems, $\langle \sigma_{yy} \rangle/E^*$ is plotted as a function of the packing fraction $\Phi$ in Fig.~\ref{fig : sigma yy}. We choose to deal with the packing fraction since it is a function of the actual shape of the particles, hence a function of the deflection of the particles boundary. The curves are found to collapse on a single curve for frictionless particles (Fig. \ref{fig : sigma yy}(a)). Strikingly, this collapse is good even at large deformation, whereas Eq.~(\ref{eqn : Hertz}) aims to describe only infinitesimal deformations. This shows that, for a given packing fraction, the compressive stress applied to a granular material with particles having a given Poisson's ratio, can be deduced from a system with particles having another given Poisson's ratio. Moreover, this collapse is less accurate as the coefficient of friction $\mu_f$ increases, but remains relevant up to small discrepancies (Fig.~\ref{fig : sigma yy}(b)). %For instance, for the packing fraction $\Phi=0.92$, the relative variation of $\langle \sigma_{yy} \rangle/E^*$ is less than 2\% for coefficients of friction $\mu_f=0.2$, and less than 23\% for coefficients of friction $\mu_f=0.8$.

At this point, it has been demonstrated that the effects of the compressibility of the particles on the global stress $\langle \sigma_{yy} \rangle$ can be accounted for by considering the packing fraction and the effective elastic modulus $E^*$. We now go further by considering the changes induced by a variation of the compressibility of the particles on the details of the stress distribution in the system. We show below that, for any fixed value of the packing fraction, the stress distribution does not depend on the Poisson's ratio of the particles, provided again that the coefficient of friction is moderate. 

\begin{figure}[!h]
\begin{center}
\includegraphics[width=0.4\textwidth]{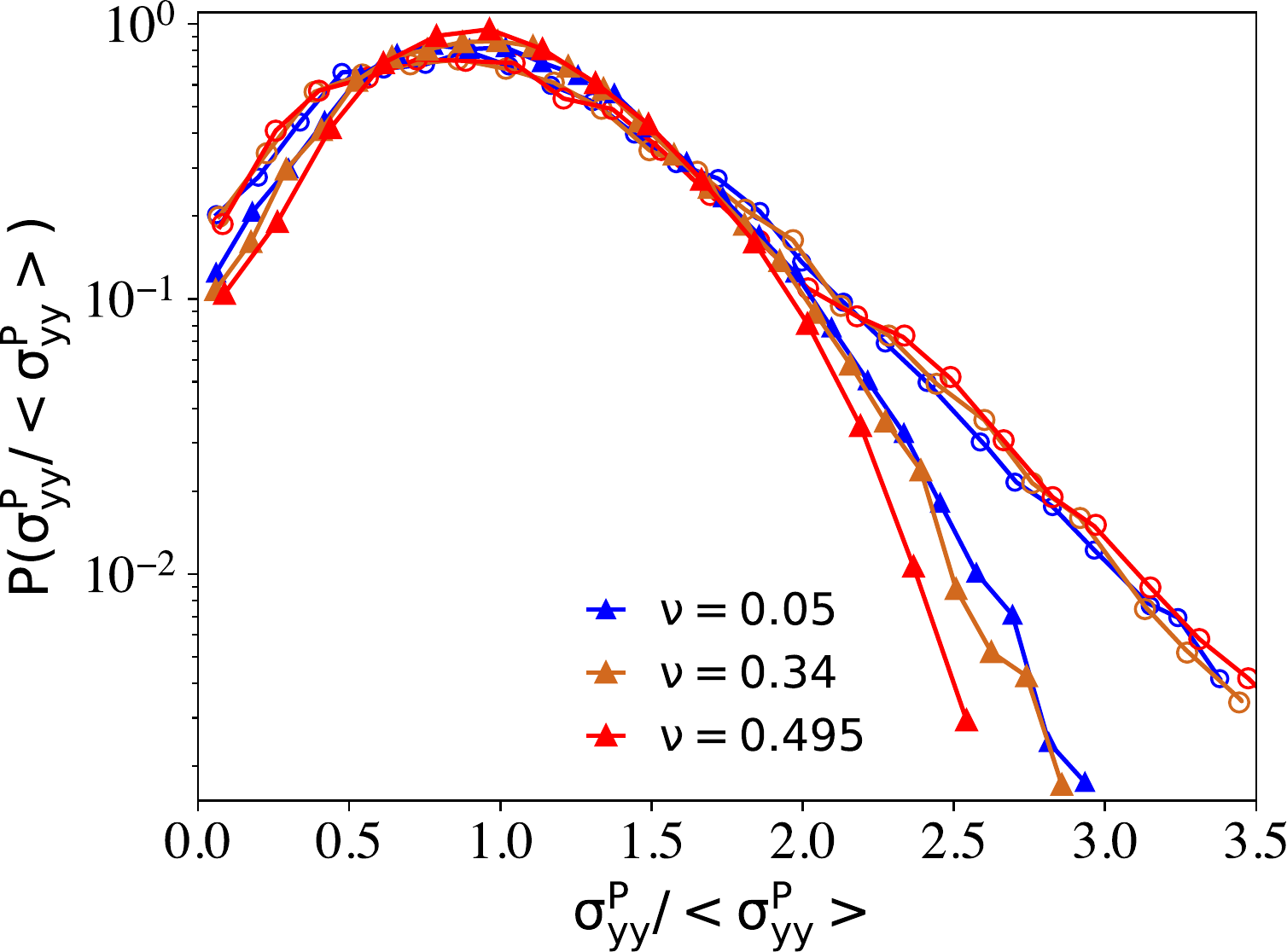}
\end{center}
\caption{Probability density function of the  mean $yy$ component of the Cauchy stress per particle for $\mu_f=0$ (triangles) and $\mu_f=0.8$ (circles), for $\nu=0.05$ (red), $\nu=0.34$ (brown) and $0.495$ (blue), and for $\Phi=0.86$.}
\label{fig : pdf}
\end{figure}

Let $\sigma^p_{yy}$ be the mean $yy$ component of the Cauchy stress of a given particle, labelled here with index $p$. The probability density function (pdf) of $\sigma^p_{yy}$ normalized by the mean value of $\sigma^p_{yy}$ is plotted in Fig.~\ref{fig : pdf} for packing fraction $\Phi=0.86$ for the three tested Poisson's ratios and for two coefficients of friction, $\mu_f=0$ and $\mu_f=0.8$. For this packing fraction, the pdf of $\sigma_{yy}^p/\langle \sigma_{yy}\rangle$ is found to be almost independent of the Poisson's ratio for the frictionless systems, while differences are more and more pronounced as the coefficient of friction increases. 
\begin{figure}[!h]
\begin{center}
  \includegraphics[width=0.4\textwidth]{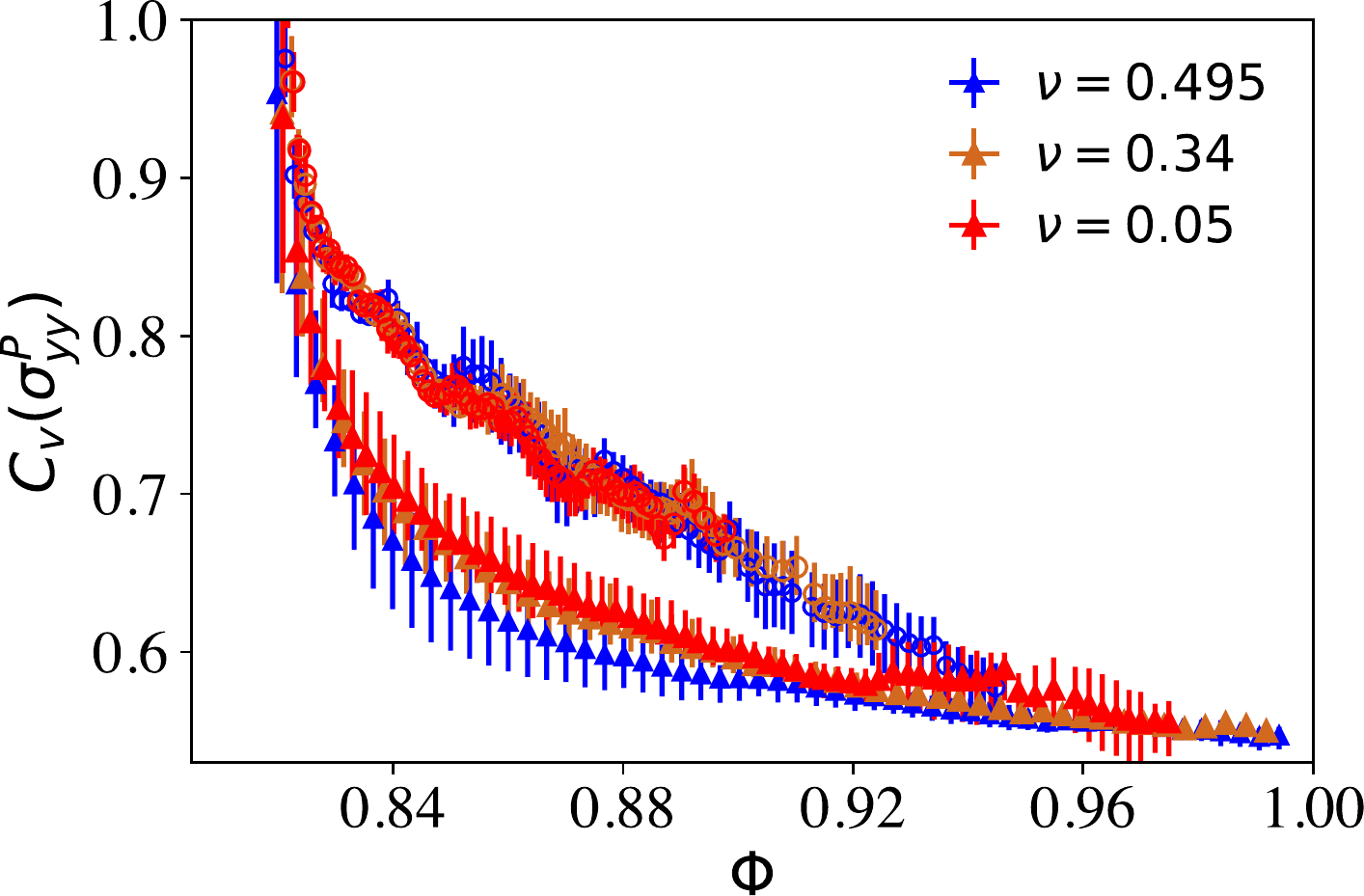}
\end{center}
\caption{Coefficient of variation of $\sigma_{yy}^p$ as a function of the packing fraction, for $\mu_f=0$ (circles) and $\mu_f=0.8$ (triangles),  for $\nu=0.495$ (blue), $\nu=0.34$ (brown) and $\nu=0.05$ (red).}
\label{fig : Cv}
\end{figure}

\begin{figure}[!h]
\begin{center}
  \includegraphics[width=0.4\textwidth]{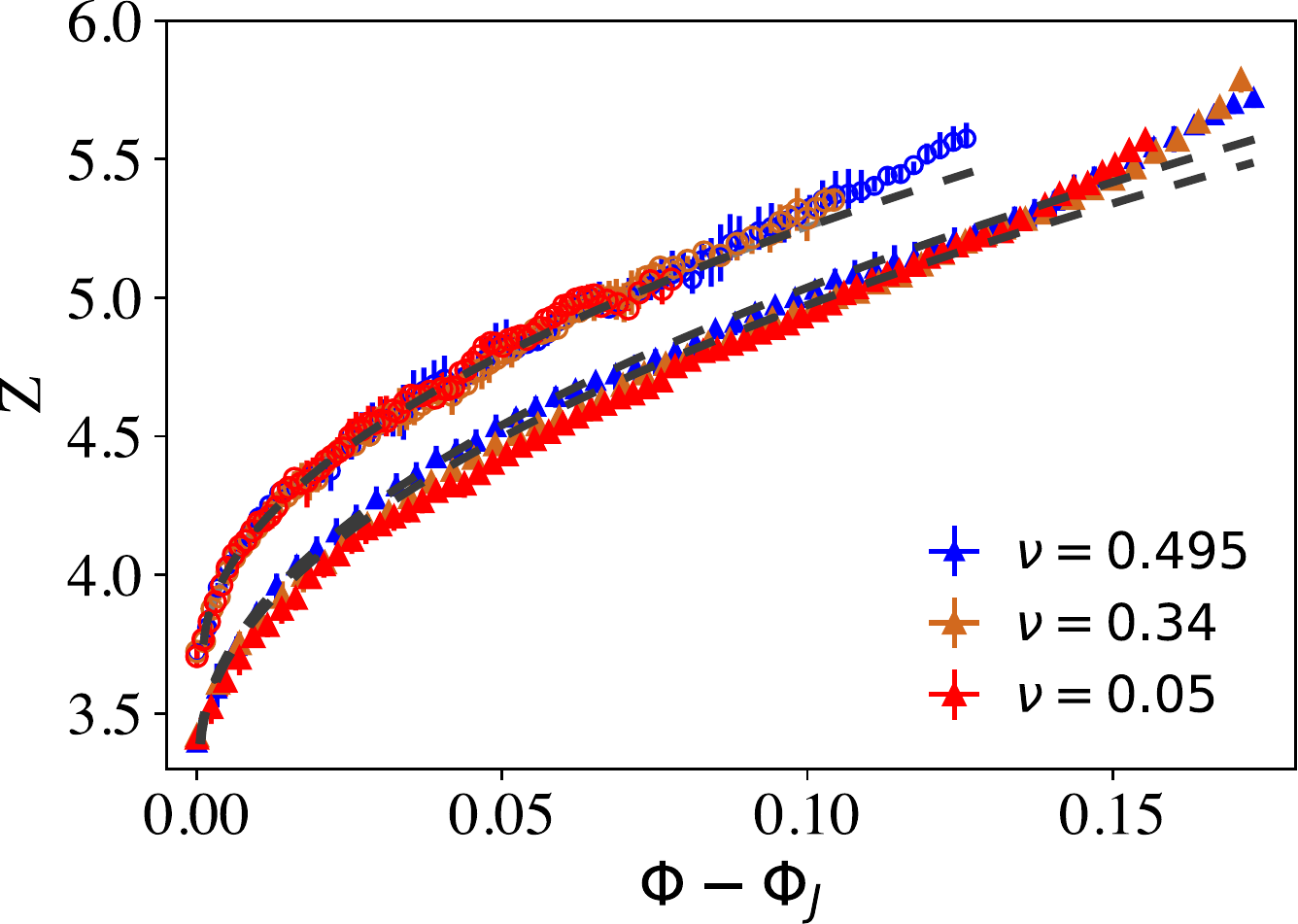}
\end{center}
\caption{Mean coordination number $Z$ as a function of the excess packing fraction $\Phi - \Phi_J$ for $\nu=0.495$ (blue), $\nu=0.34$ (brown) and $\nu=0.05$ (red), and for $\mu_f=0$ (circles) and  $\mu_f=0.8$ (triangles).}
\label{fig : Z}
\end{figure}

 The conclusion is the same for the other values of the packing fraction: Fig.~\ref{fig : Cv} shows the coefficient of variation (or the relative standard deviation) of $\sigma^p_{yy}$, defined as \cite{Hofmann2000TheEO}:
\begin{equation}
C_v(\sigma_{yy})=\frac{\langle \sigma_{yy}^2\rangle-\langle \sigma_{yy}\rangle^2 }{\langle \sigma_{yy} \rangle^2},
\end{equation}
as a function of the packing fraction. $C_v$ is calculated directly from the pdf of $\sigma^p_{yy}$.  $C_v$ decreases as the packing fraction increases beyond its critical value $\Phi_J$ for the jamming transition. It means that the system is less and less heterogeneous as the packing fraction increases, as also observed in \cite{Vu2019_soft_matter} for incompressible particles. This conclusion remains valid for all the tested values of the Poisson's ratio. Indeed, the coefficient of variation is found to be almost independent of the Poisson's ratio for frictionless particles, but small deviations appear as the inter-particles coefficient of friction increases; see Fig.~\ref{fig : Cv}. The local structure of the stress therefore is not significantly independent of the Poisson's ratio of the particles, for a given value of the packing fraction and for a given value of the coefficient of friction, provided that the coefficient of friction remains moderate. This property is shared by other structural features of the system, as the mean coordination number $Z$, defined as the average number of contact neighbors per particle. The coordination number is plotted in Fig.~\ref{fig : Z}. For frictionless system, $Z$ only depends on the packing fraction, whereas small deviations are observed while changing the Poisson's ratio in case of friction between the particles.

%%%%%%%%%%%%%%%%%%%%%%%%%%%%%%%
\section{Irreversibility} \label{sec : irreversibility}

\begin{figure}[!h]
\begin{center}
  \includegraphics[width=0.4\textwidth]{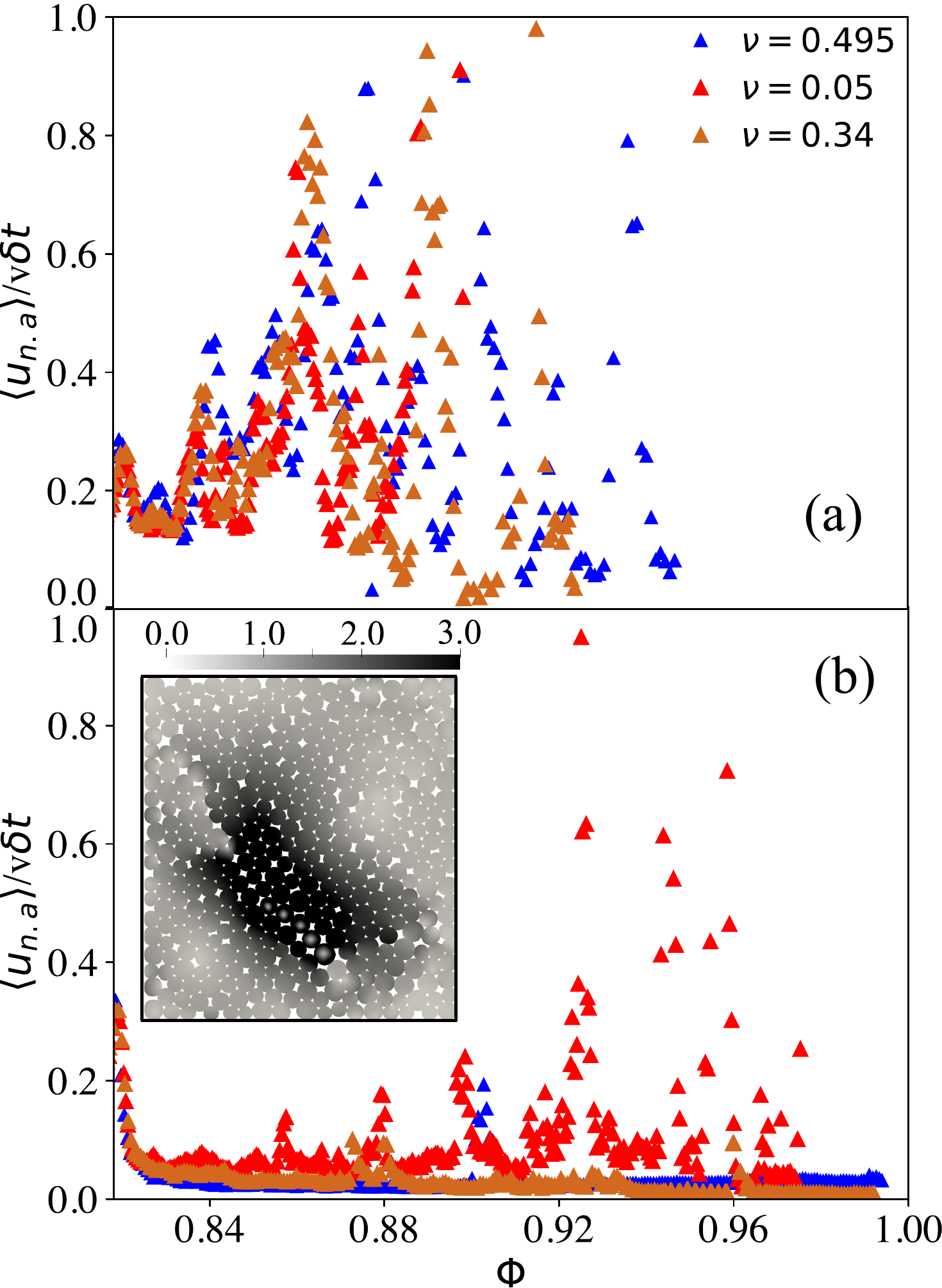}
\end{center}
\caption{Mean non-affine displacement computed for systems with $\mu_f=0$ (a) and $\mu_f=0.8$ (b). Inset:  Map of the magnitude of the non-affine displacement in a system for $\Phi=0.91$, with $\mu_f=0.2$ and $\nu=0.34$. The gray scale is for the magnitude of $ u_{n.a (\mathbf{r})} / v \delta t$.}
\label{fig : non affine}
\end{figure}

The non-affine displacement field $u_{n.a}(\mathbf{r})$ \cite{Lemaitre2006} is measured by computing the displacement at each nodes of the system during an increment $v \delta t=\SI{1.14e-4}{\meter}$ of the position of the mobile wall (with $\delta t$ as the elapsed time of the interval). The mean non-affine displacement $\langle u_{n.a}\rangle$ spatially averaged and normalized by $v \delta t$,  is plotted in Fig.~\ref{fig : non affine} as a function of the packing fraction $\Phi$ for systems with $\mu_f=0$ or $\mu_f=0.8$, and for different values of the Poisson's ratio.
%$u_{n.a}(\mathbf{r})$ spatially averaged over the whole system and normalized by $v \delta t$,  is plotted in Fig.~\ref{fig : non affine} as a function of the packing fraction $\Phi$ for systems with $\mu_f=0$ or $\mu_f=0.8$, and for different values of the Poisson's ratio.

The peaks of $\langle u_{n.a}\rangle/v \delta t$ appearing in these plots indicate the steps in the compression for which, somewhere in the system, local non-affine displacements occur; see inset of  Fig.~\ref{fig : non affine} and \cite{supplement}. For the frictionless systems, no significant differences are observed by changing the Poisson's ratio: neither the frequency nor the characteristic amplitude of the peaks seem to depend on the particle compressibility. This is in contrast with the systems with $\mu_f=0.8$, for which the peaks of the non-affine displacements are more numerous as the particles are more and more compressible. One concludes that in frictionless systems, the non-affine displacement does not significantly depend on the Poisson's ratio of the particles, whereas the non-affine properties of the systems depend to the Poisson's ratio in frictional systems. This point is consistent with the results of Section~\ref{sec : stress}: For frictionless particles, the systems behave equivalently for any fixed packing fraction, whereas for frictional particles, differences in the stresses are observed. One can then infer that these differences in the stress are related to local rearrangements occurring in the system.

%\begin{figure}[!h]
%  \begin{center}
%    \includegraphics[width=0.4\textwidth]{0_Analyse_Poisson_phi_LC.pdf}
%\includegraphics[width=0.4\textwidth]{08_Poisson_phi_LC.pdf}
%\end{center}
%\caption{Contacts length normalized with the total perimeter for $\mu_f=0$ (A) and $\mu_f=0.8$ (B). \colorbox{magenta}{$P$ est-il bien le perimetre actuel ?}  \colorbox{magenta}{Supprimer $\nu=0.34$ ?} \colorbox{magenta}{Ne peut on pas tracer toutes les} \colorbox{magenta}{courbes sur le meme graph ?}}
%\label{fig : LC}
%\end{figure}

To conclude, the slight but measurable differences observed in the stress for a given value of the packing fraction, upon a change of the Poisson's ratio of the particles, are related to the plastic events, whose main features do depend on the Poisson's ratio in frictional systems, but not in frictionless systems. 

\section{Discussion and conclusion} \label{sec : conclusion}
Effects of the particles compressibility during uni-axial compaction of two-dimensional athermal granular assemblies composed of neo-Hookean elastic particles have been investigated using simulations based on the FEM-CD method. A wide range of packing fractions has been explored, including far beyond the jamming transition, and different values of the inter-particles coefficient of friction have been tested. The compressive stress applied to the granular material, the distribution of the stress inside the system, the mean coordination number, as well as the non-affine displacements have been computed. Their variations as a function of the compressive strain ($\varepsilon$) strongly depend on the Poisson's ratio of the particles. Plotting now these quantities as a function of the actual packing fraction, the corresponding curves do not depend anymore on the Poisson's ratio of the particles for frictionless particles, showing that the particle compressibility does not bring more complexity in these granular systems. The adimensionalization leading to a master curve for the stress has been inspired by Hertz's theory. Even if this theory is established for infinitesimal strains, we have shown that this reduction is indeed relevant even in granular materials in which the particles undergo large deformations.

In frictional systems, we have found that plastic events depend on the Poisson's ratio at fixed packing fraction and fixed inter-particles coefficient of friction. Accordingly, the dimensionless stress plotted as a function of the packing fraction does not perfectly follow a master curve, the discrepancies being more marked as the coefficient of friction and the packing fraction are higher. Indeed, inter-particles friction reduces the ability of the particles to rearrange, compared with the case of frictionless particles. Whereas in frictionless systems, plastic events can result from particles sliding, rearrangements in frictional systems require a higher deformation of the particles so that they can sneak between each other, and hence depend on the compressibility of the particles. This picture gives hints to explain why irreversible features depend on the compressibility in frictional systems.

%The effects of the particles compressibility have been demonstrated in the uni-axial geometry of deformation. Other geometries, as pure shear, have now to be investigated in order to extend these results to more general deformations of these granular systems.
%In the systems considered previously, all the neo-Hookean particles have the same Poisson ratio (and also the same Young modulus), hence identical effective elastic modulus $E^*$. Mixtures of particles with different elastic properties would be interesting to consider. Would the mean value of $\mu$ or $E^*$ lead to a master curve ? What would be the effect of their distribution/dispersity ?

%Note that standard materials with positive Poisson ratio have been considered here. The case of auxetics particles, with negative Poisson ratio, has not been investigated. This would require to modify the non linear description of the constitutive law \cite{Ciambella2014}.\\

This paper deals with the effect of an unique characteristic in the constitutive law of the particles, the compressibility of neo-Hookean particles. In order to reflect the wide range of the mechanical properties of the particles met in real granular systems, a more comprehensive study would be suitable, with the goal to determine whether other particles properties (as the maximum extension in the Gent Model \cite{Gent1996} and the parameters defining the Mooney-Rivlin model \cite{Rivlin1948}  for hyperelastic particles,  or plasticity of the particles), or a polydispersity in these material properties, can be accounted for in terms of adimensionalized variables. 

\vglue 0.05\textwidth
\noindent {\bf Conflict of Interest:} The authors declare that they have no conflict of interest.

%\bibliography{Biblio}   

\begin{thebibliography}{10}

\bibitem{Hinrichsen2004}
H.~Hinrichsen and E.D. Wolf.
\newblock {\em The Physics of Granular Media}.
\newblock Wiley-VCH Verlag, 2004.

\bibitem{Radjai1996}
F.~Radjai, M.~Jean, JJ~Moreau, and S.~Roux.
\newblock Force distributions in dense two-dimensional granular systems.
\newblock {\em Phys. Rev. Lett.}, 77:274, 1996.

\bibitem{Majmudar2007}
TS. Majmudar, M.~Sperl, S.~Luding, and RP. Behringer.
\newblock Jamming transition in granular systems.
\newblock {\em Phys. Rev. Lett.}, 98:058001, 2007.

\bibitem{Kabla2009}
A.J. Kabla and T.J. Senden.
\newblock Dilatancy in slow granular flows.
\newblock {\em Phys. Rev. Lett.}, 102:228301, 2009.

\bibitem{Herminghaus2005}
S.~Herminghaus.
\newblock Dynamics of wet granular matter.
\newblock {\em Advances in Physics}, 54:221--261, 2005.

\bibitem{Luding2008}
S.~Luding.
\newblock Cohesive, frictional powders: contact models for tension.
\newblock {\em Granular Matter}, 10:235--246, 2008.

\bibitem{MiDi2004}
GDR. MiDi.
\newblock Friction enhances elasticity in granular solids.
\newblock {\em European Physical Journal E}, 14:341--365, 2004.

\bibitem{Goldenberg2005}
C.~Goldenberg and I.~Goldhirsch.
\newblock Friction enhances elasticity in granular solids.
\newblock {\em Nature}, 435:188--191, 2005.

\bibitem{Azema2010}
E.~Az\'ema and Radjai.
\newblock Stress-strain behavior and geometrical properties of packings of
  elongated particles.
\newblock {\em Phys. Rev. E}, 81:051304, 2010.

\bibitem{Shaebani2012}
MR. Shaebani, M.~Madadi, S.~Liding, and DE. Wolf.
\newblock Influence of polydispersity on micromechanics of granular materials.
\newblock {\em Phys. Rev. E}, 85:011301, 2012.

\bibitem{Radjai2015}
D-H Nguyen, E.~Az\'ema, P.~Sornay, and Radjai.
\newblock Effects of shape and size polydispersity on strength properties of
  granular materials.
\newblock {\em Phys. Rev. E}, 91:032203, 2015.

\bibitem{Vu2019_pre}
T.L. Vu, J.~Bar\'es, S.~Mora, and S.~Nezamabadi.
\newblock Numerical simulations of the compaction of assemblies of rubberlike
  particles: A quantitative comparison with experiments.
\newblock {\em Phys. Rev. E}, 99:062903, 2019.

\bibitem{Nezamabadi2017}
S.~Nezamabadi, T.~H. Nguyen, J.~Y. Delenne, and F.~Radjai.
\newblock Modeling soft granular materials.
\newblock {\em Granular Matter}, 19:8, 2017.

\bibitem{Radjai2018}
D-H Nguyen, E.~Az\'ema, P.~Sornay, and Radjai.
\newblock Rheology of granular materials composed of crushable particles.
\newblock {\em Eur. Phys. J. E}, 41:50, 2018.

\bibitem{OHern2002}
C.S. O'Hern, A.L. Langer, A.J. Liu, and S.R. Nagel.
\newblock Random packings of frictionless particles.
\newblock {\em Physical Review Letters}, 88:075507, 2002.

\bibitem{Kabla2012}
A.~Kabla.
\newblock Collective cell migration: leadership, invasion and segregation.
\newblock {\em J. R. Soc. Interface}, 9:3268--3278, 2012.

\bibitem{Lepesant2013}
P.~Lepesant, C.~Boher, Y.~Berthier, and F.~R\'ezai-Aria.
\newblock A phenomenological model of the third body particles circulation in a
  high temperature contact.
\newblock {\em Wear}, 298-299:127--134, 2013.

\bibitem{Lorenzo2013}
G.~Lorenzo, N.~Zartizky, and A.~Califano.
\newblock Rheological analysis of emulsion-filled gels based on high acyl
  gellan gum.
\newblock {\em Food Hydrocoll.}, 30:672--680, 2013.

\bibitem{Menut2012}
P.~Menut, S.~Seiffert, J.~Sprakel, and D.~Weitz.
\newblock Does size matter? elasticity of compressed suspensions of colloidal-
  and granular-scale microgels.
\newblock {\em Soft Matter}, 8:156--164, 2012.

\bibitem{Ogden1984}
R.W. Ogden.
\newblock {\em Non-Linear Elastic Deformations}.
\newblock Ellis Horwood Limited, Chichester, 1984.

\bibitem{Nezamabadi2011}
S.~Nezamabadi, H.~Zahrouni, and J.~Yvonnet.
\newblock Solving hyperelastic material problems by asymptotic numerical
  method.
\newblock {\em Computational mechanics}, 47:77--92, 2011.

\bibitem{Vu2019_ExpMech}
TL. Vu, J.~Bar{\'e}s, S.~Mora, and S.~Nezamabadi.
\newblock Deformation field in diametrically loaded soft cylinders.
\newblock {\em Experimental Mechanics}, 59:453--467, 2019.

\bibitem{Vu2017}
Thi~Lo Vu, Saeid Nezamabadi, Jonathan Bar{\'e}s, and Serge Mora.
\newblock Analysis of dense packing of highly deformed grains.
\newblock {\em EPJ Web of Conferences}, 140:15031, 2017.

\bibitem{Vu2019_soft_matter}
T.L. Vu, S.~Nezamabadi, and S.~Mora.
\newblock Compaction of elastic granular materials : inter-particles friction
  effects and plastic events.
\newblock {\em Soft Matter}, 2020.

\bibitem{lmgc90}
LMGC90.
\newblock Lmgc90.
\newblock
  \url{https://git-xen.lmgc.univ-montp2.fr/lmgc90/lmgc90\_user/wikis/home},
  2018.

\bibitem{supplement}
See movie in Supplementary material showing simulations of compactions of 3
  granular systems with $\nu=0.495$, $\nu=0.34$ and $\nu=0.05$ and $\mu_f=0.2$.
  Initial packing fractions are $\Phi=81\%$. Final packing fractions are
  $\Phi=97\%$.

\bibitem{Johnson1985}
K.L. Johnson.
\newblock {\em Contact Mechanics}.
\newblock Cambridge University Press, Cambridge, 1985.

\bibitem{Hofmann2000TheEO}
Werner Hofmann, Bahman Asgharian, R~Bergmann, Satish Anjilvel, and Frederick~W
  Miller.
\newblock The effect of heterogeneity of lung structure on particle deposition
  in the rat lung.
\newblock {\em Toxicological sciences : an official journal of the Society of
  Toxicology}, 53 2:430--7, 2000.

\bibitem{Lemaitre2006}
C.~Maloney and A.~Lemaitre.
\newblock Amorphous systems in athermal, quasistatic shear.
\newblock {\em Phys. Rev. E}, 74:016118, 2006.

\bibitem{Gent1996}
A.N. Gent.
\newblock A new constitutive relation for rubber.
\newblock {\em Rub. Chem. Tech.}, 69:59--61, 1996.

\bibitem{Rivlin1948}
R.S. Rivlin.
\newblock Large elastic deformations of isotropic materials. iv. further
  developments of the general theory.
\newblock {\em Philosophical Transactions of the Royal Society of London.
  Series A, Mathematical and Physical Sciences}, 241:379--397, 1948.

\end{thebibliography}

\end{document}